\documentclass[twocolumn,times]{aastex62}
\usepackage{amsmath}
\usepackage{amssymb}
\usepackage{lineno}
\allowdisplaybreaks[4]

\usepackage{color}

\shorttitle{Formation of retrograde hot Jupiter triggered by close encounter}
\shortauthors{Wenshuai Liu}

\begin{document}

\title{\Large{\textbf{Formation of Retrograde Hot Jupiter Triggered by Close Encounter with Tidal Circularization}}}

\correspondingauthor{Wenshuai Liu}
\email{674602871@qq.com}

\author{Wenshuai Liu}
\affiliation{School of Physics, Henan Normal University, Xinxiang 453007, China}



\begin{abstract}
 A new mechanism is proposed to account for the formation of retrograde hot Jupiter in coplanar star-planet system via close encounter between a Jupiter mass planet and a brown dwarf mass planet. After long timescale scattering between several Jupiter mass planets with inner orbits, the remaining planets still rotating around the star could have large semimajor axis with large eccentricity. If there exists a brown dwarf mass planet in distant orbit around the star, planetary encounter may happen. After encounter, the Jupiter mass planet may rotate around the star in a retrograde orbit with extremely large eccentricity and the periastron can reach about 0.01 AU, which means that, within the first several orbits around the star, tidal interaction from the star can shrink the semimajor axis of the planet quickly. Thus, the Jupiter mass planet is isolated from the brown dwarf mass planet due to the quick decrease of its apastron distance and eventually evolves into a retrograde hot Jupiter.
\end{abstract}

\keywords{planet-star interactions -- planets and satellites: dynamical evolution and stability -- planets and
satellites: formation}


\section{Introduction}

Hundreds of hot Jupiters (HJs) have been detected around their host main sequence stars. Due to the fact that HJs could not form at their current locations where the gas temperature is high and disk mass is low, a common accepted idea is that they first form at lager distance far away from their host stars and then migrate to the current place through angular momentum exchange with the protoplanetary disk \citep{1,2} or by tidally circularizing a highly eccentric orbit. For the former scenario, HJs should have small eccentricity and orbital inclination because disks tend to damp eccentricity and inclination of the planet's orbit. However, despite the observed HJs with features which can be explained by the former scenario, there are some HJs with high eccentricity and large orbital inclination. To explain such features, some proposed mechanisms accounting for the origin of the high eccentricity include planet-planet scattering \citep{3,4,5,6}, Kozai-Lidov (KL) mechanism \citep{7,8,9,10,11,16,12,13} and secular chaos between planets \citep{14,15}.

Based on the Rossiter-McLaughlin (RM) effect, the observed sky-projected angle between the spin of the central star and the orbit of the planet, or the projected spin-orbit angle for short, $\lambda$, can be larger that $90^o$, meaning that such HJs are in the retrograde orbit. Mechanisms that can account for such retrograde HJs include KL mechanism \citep{11}, secular chaos between planets \citep{15} and coplanar high eccentricity migration \citep{17,18,19,21}.

Here, I propose a new mechanism of the formation of retrograde hot Jupiter in coplanar star-planet system, which is a natural result of close encounter followed by tidal circularization after some dynamical processes producing high eccentricity planetary orbit.

The model is described in Section 2 and the discussion is given in Section 3.

\section{The Model}
After dynamical processes (planet-planet scattering, for example) which produce Jupiter mass planet with high eccentricity, planetary encounter could happen if there is a brown dwarf mass planet in the distant orbit around the same host star and if the two orbits cross.

\begin{figure}
            \includegraphics[width=0.5\textwidth]{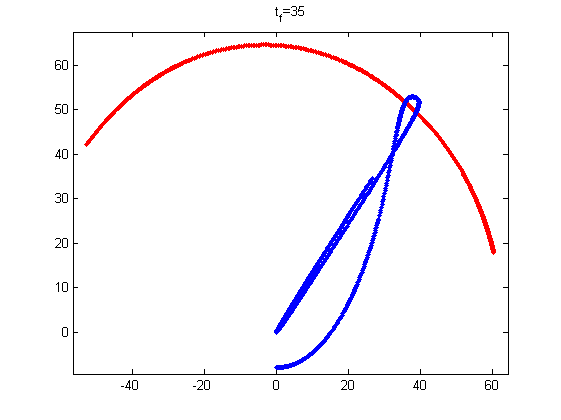}
\caption{ Blue and red curve show the trajectory of the Jupiter mass planet and the brown dwarf mass planet, respectively.}
\label{fig:figure1}
\end{figure}

\begin{figure}
            \includegraphics[width=0.5\textwidth]{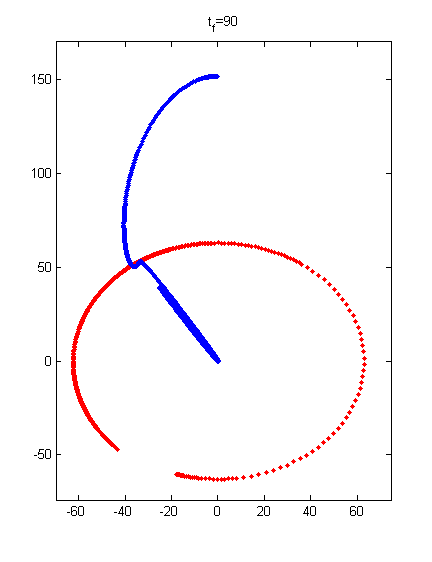}
\caption{ Blue and red curve show the trajectory of the Jupiter mass planet and the brown dwarf mass planet, respectively.}
\label{fig:figure1}
\end{figure}

Three-body gravitational interaction are adopted to simulate the encounter which produce Jupiter mass planet with retrograde orbit. We set the mass of the star $M=M_\odot$, the mass of the brown dwarf mass planet $m_b=0.03M_\odot$ and the mass of the Jupiter mass planet $m_J=0.001M_\odot$, and use units $G=1$, $M=1$ and $1AU=1$. The $x-y$ coordinate of the host star is $(0,0)$ with velocity (0,0) while the nearly circular orbital radius of the brown dwarf mass planet is $a=63$AU with initial position $\{a*\sin(\frac{73.55\pi}{180}),a*\cos(\frac{73.55\pi}{180})\}$ and initial velocity $\{\sqrt{\frac{G(M+m_b)}{a}}\cos(\pi-\frac{73.55\pi}{180}),\sqrt{\frac{G(M+m_b)}{a}}\sin(\pi-\frac{73.55\pi}{180})\}$. The semimajor axis of the Jupiter mass planet is $a_J=80$AU with eccentricity of $e=0.9$ (according to the results in \cite{4}). The initial position of the Jupiter mass planet is at periastron with coordiantes $\{0,-8.0\}$ with initial velocity $\{15.411,0\}$. With these initial conditions and only Newtonian gravitation, the encounter could occur and we run the simulation until $t_f=35$. The results in Figure 1 show clearly that the initially prograde Jupiter mass planet changes to the retrograde direction around the star after encounter with the brown dwarf mass planet.

During encounter, the smallest distance between the two planets is about 1AU, meaning that tidal dissipation can be neglected. After encounter, the Jupiter mass planet could approach the star at the closest distance of about $0.0132$AU which is outside the Roche limit according to $d=R(2\frac{M}{m_J})^\frac{1}{3}$ where we set $R$ twice the Jupiter's radius. However, tidal interaction with the star can reduce the velocity of the Jupiter mass planet obviously at such small distance. Here, similar as \citep{4}, the relative velocity of the Jupiter mass planet with respect to the star is changed discontinuously at the pericenter passage described as

\begin{equation}
{\bf v}^{\prime}= \sqrt{2\Delta E_{\rm tide} +v^2}\frac{{\bf v}}{v},
\label{eq:vchange}
\end{equation}
where $\Delta E_{\rm tide}$ is given by
\begin{equation}
    \Delta E_{tide}\  \sim - \frac{16\sqrt{2}}{15} {\widetilde{w_0}}^{3}{\widetilde{Q}}^{2} \xi \exp\left(- \frac{4\sqrt{2}}{3}\widetilde{w_0}\xi \right)\frac{Gm^{2}}{R}\ ,
    \label{eq:e_tide}
\end{equation}
in which $\widetilde{w_0}\simeq 0.53(R/R_{J})+0.68$ and $\widetilde{Q}\simeq -0.12(R/R_{J})+0.68$ for the Jupiter mass planet, $R_{J}$ is Jupiter's radius, and $\xi = (mq^{3})^{1/2}(MR^{3})^{-1/2}$. $q$ is the periastron distance. $m$ and $R$ are the mass and radius of the Jupiter mass planet, respectively.

With $m=m_J$ and $R=2R_J$, after the first pericenter passage, the relative velocity of the Jupiter mass planet with respect to the star is changed to $v^{\prime}= \sqrt{2\Delta E_{\rm tide} +v^2}$ where $v$ is the velocity at periastron in the absence of tidal dissipation. The corresponding apastron of the Jupiter mass planet changes from about 66.9AU to about 38.2AU. This means that the Jupiter mass planet is isolated from the brown dwarf mass planet. After six orbits of the Jupiter mass planet around the star, the apastron decreases to about 12.14AU. When effect of general relativity is taken into account, the planet's orbit will precess. After several tens orbits, the eccentricity becomes moderate, and the evolutions of the semimajor axis and the eccentricity are given as follows \citep{24,22,23,20}

\begin{eqnarray}
\frac{da}{dt} && =\frac{6P_1}{M_{*}a^{4}}
\frac{\left[(1-e^{2})^{3/2}f_{2}(e^{2})\omega_{*}\cos\epsilon_*-f_{1}(e^{2})n \right]}{(1-e^{2})^{15/2}}
\nonumber \\
 && +  \frac{6P_2}{M_{p}a^{4}}
\frac{\left[(1-e^{2})^{3/2}f_{2}(e^{2})\omega_{p}-f_{1}(e^{2})n \right]}{(1-e^{2})^{15/2}}\\
\nonumber \\
\frac{de}{dt} && =\frac{27P_1}{M_{*}a^{5}}
\frac{\left[\frac{11}{18}(1-e^{2})^{3/2}f_{4}(e^{2})\omega_{*}\cos\epsilon_*
-f_{3}(e^{2})n \right]}{e^{-1}(1-e^{2})^{13/2}}
\nonumber \\
 && +  \frac{27P_2}{M_{p}a^{5}}
\frac{\left[\frac{11}{18}(1-e^{2})^{3/2}f_{4}(e^{2})\omega_{p}-f_{3}(e^{2})n \right]}{e^{-1}(1-e^{2})^{13/2}}
\end{eqnarray}
where $P_1=k_{2,*}\Delta t_*nM_{p}R_{*}^{5}$, $P_2=k_{2,p}\Delta t_pnM_{*}R_{p}^{5}$ and the details can be found in works \citep{22,23,24,20}.

When the initial position of the Jupiter mass planet changes to the apastron, the position and velocity are $\{0,152\}$ and $\{-0.8111,0\}$, respectively. The initial position of the brown dwarf mass planet is $\{a*\sin(\frac{196.6\pi}{180}),a*\cos(\frac{196.6\pi}{180})\}$ in order to have close encounter. The results in Figure 2 show similar property as that in Figure 1.

The two simulations above present the detailed dynamical close encounter producing the retrograde orbit. When it comes to a planetary system with three or more well-spaced, coplanar circular planets, it shows from below that retrograde orbit could also occur.

With the initial conditions that four Jupiter mass planets with orbital radius of $a=3,4,5,6AU$ and with coplanar circular orbit around the host star of the mass of the sun, our preliminary results from the N-body simulations show that, during the dynamical evolution, retrograde orbit induced by close encounter occurs.
\section{Discussion}
This work show that retrograde hot Jupiters could form as a natural result of close encounter with a brown dwarf mass planet followed by tidal circularization. Thus, the mechanism proposed here provides a potential way of forming retrograde hot Jupiter. Here, two typical sets of initial parameters are adopted in order to produce the close encounter, it should be noted that, depending on the initial condition, close encounter produces not only retrograde Jupiter, but also prograde one. In some extreme conditions, the Jupiter mass planet will hit the star in the retrograde direction after encounter with the brown mass planet, resulting that the host star will spin in the retrograde direction. This work adopt a coplanar star-planet system, it should be noted that similar results could also be produced when the inclination of planets' orbit is small. To cover the initial parameter space, comprehensive parameter survey will be conducted to study the efficiency of this mechanism in the near future. Although Jupiter mass planet is considered in this work, we may expect that similar results could happen if a stellar mass black hole orbiting around a supermassive black hole encounters an intermediate mass black hole.


\end{document}